\documentclass[conference]{IEEEtran}
\usepackage[T1]{fontenc}
\usepackage[latin9]{inputenc}
\usepackage{float}
\usepackage{amsthm}
\usepackage{amsmath}
\usepackage{amssymb}
\usepackage{stmaryrd}
\usepackage{graphicx}

\makeatletter
\theoremstyle{plain}

\theoremstyle{definition}
\newtheorem{defn}{\protect\definitionname}
\theoremstyle{remark}
\newtheorem{rem}{\protect\remarkname}
\theoremstyle{plain}
\newtheorem{lem}{\protect\lemmaname}
\theoremstyle{plain}
\newtheorem{prop}{\protect\propositionname}

\usepackage{textcomp}
\usepackage{dsfont}\usepackage{cite}
\usepackage{color}

\providecommand{\definitionname}{Definition}
\providecommand{\lemmaname}{Lemma}
\providecommand{\propositionname}{Proposition}
\providecommand{\remarkname}{Remark}

\IEEEoverridecommandlockouts

\begin{document}

\title{Lossy Compression with Privacy Constraints: Optimality of Polar Codes}

\author{Farshid Mokhtarinezhad, J{ö}rg Kliewer, Osvaldo Simeone \\[1ex]
Helen and John C.~Hartmann Department of Electrical and Computer
Engineering\\
 New Jersey Institute of Technology\\
 Newark, New Jersey 07102-1982\\
 Email: \{fm86, jkliewer, simeone\}@njit.edu %
\thanks{This work was supported in part by NSF grant CCF-1439465.%
}}
\maketitle
\begin{abstract}
A lossy source coding problem with privacy constraint is studied
in which two correlated discrete sources $X$ and $Y$ are compressed
into a reconstruction $\hat{X}$ with some prescribed distortion $D$.
In addition, a privacy constraint is specified as the equivocation
between the lossy reconstruction $\hat{X}$ and $Y$. This models
the situation where a certain amount of source information from one
user is provided as utility (given by the fidelity of its reconstruction)
to another user or the public, while some other correlated part of
the source information $Y$ must be kept private. In this work, we
show that polar codes are able, possibly with the aid of time sharing,
to achieve any point in the optimal rate-distortion-equivocation region
identified by Yamamoto, thus providing a constructive scheme that
obtains the optimal tradeoff between utility and privacy in this framework.
\end{abstract}

\section{Introduction}

An important consequence of the ubiquitous growth of modern information
technology is that an increasing amount of private information is
shared between different organizations and/or users. This entails
a tension between privacy and utility in the sense that disclosing
data provides useful information to the receiving entity, while at
the same time posing the danger of leaking  private information.
Examples for such a tension can be found in many real-life systems,
e.g., in social networks, smart grids, or databases.

The tradeoff between utility and privacy has been the subject of several
recent works as surveyed in \cite{STRVD13}. A simple information-theoretic
model to analyze this tradeoff is the lossy source coding problem
introduced by \cite{Yam83}, where utility is measured by the reconstruction
fidelity and privacy by an equivocation (i.e., conditional entropy).
Reference \cite{Yam83} shows that (vector) quantization, as realized
by means of random coding, is optimal in the sense that is achieves
any point in the rate-distortion-equivocation region. Several subsequent
works \cite{AS00,DSNS08,SRP13,MSFM14,KOS14} have addressed related
problems in which the introduction of distortion is used to disguise
private information. For example, \cite{SRP13} focuses on database
privacy in which only certain entries of a database are to be published.
Further, the authors in \cite{KOS14} generalize the result in \cite{Yam83}
to the case with side information at the decoder. While all these
works consider achievability based on random coding, here we focus
on the general setup in \cite{Yam83} and provide a constructive coding
scheme based on polar codes which achieves the optimal rate-utility-privacy
trade-off.

Polar codes, as first proposed in \cite{Ari09}, are binary block
codes which achieve the capacity of a binary symmetric memoryless
channel with efficient encoding and decoding algorithms. The key property
of these codes is that they yield virtual channels which either asymptotically
converge to an error-free or a completely noisy channel, such that
the fraction of asymptotically error-free channels approaches the
symmetric capacity of original channel. Polar codes have been generalized
to both asymmetric channels \cite{HY13,MHSU14} and arbitrary alphabets
\cite{Sas12}. Moreover, polar codes have been shown to achieve the
rate-distortion bound for symmetric binary sources \cite{KU10} and
asymmetric binary sources under Hamming distortion in \cite{HY13,ELKLJB14}.

In the following, we show that, for the framework in \cite{Yam83}
under the assumption of prime source alphabets, polar codes are able,
possibly with the aid of time sharing, to achieve any point in the
optimal rate-distortion-equivocation region. To the best of our knowledge,
this is the first constructive scheme that is provably optimal in
terms of the achievable tradeoff between rate, utility, and privacy.



\textit{Notation: }An upper case letter $A$ denotes a random variable
and $a$ denotes its realization. We let $A^{i}$ denote the random
vector $(A_{1},...,A_{i})$. For any set ${\cal S}$, $\left|{\cal S}\right|$
denotes its cardinality and $A_{{\cal S}}$ denotes the vector $(A_{i_{1}},...,A_{i_{\left|{\cal S}\right|}})$.

\begin{figure}[tb]
\centering\includegraphics[scale=0.5]{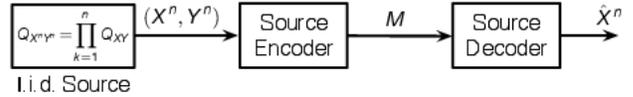} 
 \protect\caption{Illustration of the problem of lossy source coding with privacy constraints
\cite{Yam83}, in which the privacy is measured by the leakage $H(Y^{n}|M)/n$
and the utility by the fidelity as gauged with respect to the expected
distortion $\mathbb{E}(d(\hat{X}^{n},X^{n}))/n$. }

\label{fig:system}
\end{figure}

\section{System Model and Preliminary Results}

We consider the lossy source coding set-up studied in \cite{Yam83}
and depicted in Fig.~\ref{fig:system}, in which the encoder wishes
to communicate a source sequence $X^{n}$ within some distortion to
the decoder, while keeping the receiver's knowledge about a correlated
sequence $Y^{n}$, also available to the encoder, below some tolerated
level. The sources $X^{n}\in\mathcal{X}^{n}$ and $Y^{n}\in\mathcal{Y}^{n}$
take values in discrete alphabets $\mathcal{X}$ and $\mathcal{Y}$,
and are memoryless with joint distribution $Q_{X^{n}Y^{n}}(x^{n},y^{n})=\prod_{i=1}^{n}Q_{XY}(x_{i},y_{i})$
for some joint pmf $Q_{XY}(x,y)$. Encoding of these pairs leads to
an index $M=m$ with $m\in\{1,2,\dots,\lfloor q^{nR}\rfloor\}$, where
$n$ is the blocklength and $q$ is a prime number. Finally, the reconstruction
of $X^{n}$ at the decoder is given by a sequence $\hat{X}^{n}\in\hat{\mathcal{X}}^{n}$,
with $\hat{\mathcal{X}}=\{0,1,...,q-1\}$. The goal in designing the
system in Fig.~\ref{fig:system} is to obtain a desired tradeoff
between the rate $R$, the expected distortion $\mathbb{E}(d(\hat{X}^{n},X^{n}))/n$,
and the information leakage $H(Y^{n}|M)/n$ about the source $Y^{n}$
that can be obtained from observing $M$. For simplicity, in the following
we will identify pmfs by their arguments only and drop any subscripts.

We now define the operation of both encoder and decoder and the notion
of the rate-distortion-equivocation region.
To this end, we introduce a standard bounded distortion metric $d:\mathcal{X}\times\hat{\mathcal{X}}\rightarrow\left[0,d_{max}\right]$,
where $d_{max}<\infty$ is the maximal distortion. 
\begin{defn}[Code]
An $\left(n,R,D,\Delta\right)$ code consists of an encoding function
that maps each sequence $(x^{n},y^{n})\in\mathcal{X}^{n}\times\mathcal{Y}^{n}$
to an index $m\left(x^{n},y^{n}\right)\in\{1,2,\dots,\lfloor q^{nR}\rfloor\}$
and a decoding function that maps each index $m$ to an estimate $\hat{x}^{n}\left(m\right)\in\hat{\mathcal{X}}^{n}$,
such that the average distortion $\frac{1}{n}\mathbb{E}(d(X^{n},\hat{X}^{n}))\triangleq\frac{1}{n}\sum_{i=1}^{n}\mathbb{E}(d(X_{i},\hat{X}_{i}))$
satisfies the inequality
\begin{equation}
\frac{1}{n}\mathbb{E}(d(X^{n},\hat{X}^{n}))\leq D,
\end{equation}
and the equivocation rate guarantees the inequality%
\footnote{All the entropies will be computed with base $q$ logarithms and all
summations are done modulo $q$.%
}
\begin{equation}
\frac{1}{n}H(Y^{n}|M)\geq\Delta.
\end{equation}

\begin{defn}[Rate-distortion-equi\-vocation region]
A triple $\left(R,D,\Delta\right)$ is said to be achievable, if,
for any $\epsilon>0$ and $n$ sufficiently large, there exists an
$\left(n,R,D+\epsilon,\Delta-\epsilon\right)$ code. The closure of
all achievable triples $\mathcal{R}^{*}$ is referred to as the \textit{rate-distortion-equivocation
region}.
\end{defn}
\end{defn}
\begin{rem}
The distortion $D$ can be constrained without loss of generality
to lie in the interval $[0,d_{\text{max}}]$, while the equivocation
$\Delta$ may range in the interval $[H(Y|X),H(Y)]$.
\end{rem}

\subsection{Preliminaries}

\begin{lem}[\mbox{\!\!\cite{Yam83}}]
The rate-distortion-equivocation region $\mathcal{R}^{*}$ is given by the
closure of the union of all tuples $\left(R,D,\Delta\right)$ such
that the inequalities \begin{subequations}
\begin{eqnarray}
R & \leq & I(XY;\hat{X})\\
D & \leq & \mathbb{E}(d(X,\hat{X})),\\
\Delta & \geq & H(Y|\hat{X}),
\end{eqnarray}
\end{subequations}hold for some pmf $P\left(x,y,\hat{x}\right)$
that satisfies
\begin{equation}
\sum_{\hat{x}\in\hat{\mathcal{X}}}P\left(x,y,\hat{x}\right)=Q\left(x,y\right),\,\forall(x,y)\in\mathcal{X\times Y}.
\end{equation}
\end{lem}
\begin{rem}
From a pmf $P(x,y,\hat{x})$, the test channel 
\begin{equation}
W(x,y|\hat{x})=\frac{P(x,y,\hat{x})}{\sum_{(x,y)\in{\cal X}\times{\cal Y}}P(x,y,\hat{x})}
\end{equation}
can be calculated. In \cite{Yam83}, two specific binary examples
are worked out, namely a source in which the correlation between the
binary variables $X$ and $Y$ is a Z-channel, and a doubly symmetric
binary source, both under Hamming distortion. From the results in
\cite{Yam83}, it can be inferred that test channels (5) that yield
boundary points on the rate-distortion-equivocation region for the
former case are generally asymmetric, while for the latter they can
be assumed to be symmetric with no loss of optimality. We recall that
a channel $W(x,y|\hat{x})$ is said to be symmetric if there exists
a permutation $\pi(x,y)$ of the output alphabet $\mathcal{X}\times\mathcal{Y}$
such that, the identity $\pi(x,y)=\pi^{-1}(x,y)$ holds and the equality
$W(x,y|1)=W(\pi(x,y)|0)$ is satisfied for all $(x,y)\in\mathcal{X}\times\mathcal{Y}$.
\end{rem}

\section{Optimality of Polar Codes}

Let us define as $P(x,y,\hat{x})$ a pmf that achieves an operating
point of interest in the rate-distortion-equivocation region ${\cal R}^{*}$ in Lemma 1.
Let us also define as $R^{*}=I(XY;\hat{X}),D^{*}=\mathbb{E}(d(X,\hat{X}))$,
and $\Delta^{*}=H(Y|\hat{X})$ the rate, distortion and equivocation
attained under such distribution $P(x,y,\hat{x})$, respectively.
In this section, we demonstrate that polar codes can achieve any such
triple $(R^{*},D^{*},\Delta^{*})$ in ${\cal R}^{*}$. As mentioned, we focus in
the following on the case of a prime size alphabet $\hat{{\cal X}}=\{0,1,...,q-1\}$,
although extensions to alphabets of arbitrary cardinality are possible
by following \cite{Sas12}.

\subsection{Lossy source coding via polar codes}

We consider a polar coding scheme that is a variant of the approach
proposed in \cite{HY13} for asymmetric sources, which is in turn
inspired by \cite{MHSU14,ELKLJB14}, and extended to prime alphabets
by applying results of \cite{Sas12}. We fix a joint distribution
$P(x,y,\hat{x})$ that achieves a desired point $(R^{*},D^{*},\Delta^{*})$
in ${\cal R}^{*}$. To start, let us define the following joint distribution
on the set ${\cal X}^{n}\times{\cal Y}^{n}\times\hat{{\cal X}}^{n}\times{\cal U}^{n}$
where ${\cal U}=\{0,1,...,q-1\}$:
\begin{equation}
P(x^{n},y^{n},\hat{x}^{n},u^{n})=\prod_{i=1}^{n}Q(x_{i},y_{i})P(\hat{x}_{i}|x_{i},y_{i})\boldsymbol{1}\{u^{n}=\hat{x}^{n}G_{n}\},
\end{equation}
with $n=2^{k}$ for some integer $k$, $G_{n}=G^{\otimes k}$ is the
polarizing transform with $G=\left(\begin{array}{cc}
1 & 0\\
1 & 1
\end{array}\right)$, $G^{\otimes k}$ denotes the $k$-times Kronecker power, and $P(\hat{x}|x,y)=P(x,y,\hat{x})\diagup Q(x,y)$.
The distribution (6) can be interpreted as providing the target joint
distribution over variables $(X^{n},Y^{n},\hat{X}^{n})$ since, under
(6), it is easy to see that the desired distortion $D^{*}$ and equivocation
$\Delta^{*}$ are attained (see \cite{Yam83}). The challenge is to
construct a coding scheme that mimics (6) without having to transmit
a message $u^{n}$ of $n$ symbols and hence of rate $R=1$ from encoder
to decoder. Note that the matrix $G_{n}$
satisfies $G_{n}=G_{n}^{-1}$ and hence, from $u^{n}$, one can recover
$\hat{x}^{n}$ as $\hat{x}^{n}=G_{n}u^{n}$ \cite{Ari10}.

As explained in the following, the encoder maps the sources $(x^{n},y^{n})$
into a vector $u^{n}$, which is divided into two subvectors, namely
the information vector $u_{\mathcal{I}}$, indexed by the set ${\cal I}$
of size $\left|\mathcal{I}\right|=nR$ symbols and the complementary
vector $u_{\mathcal{I}^{c}}$. The information vector $u_{{\cal I}}$
constitutes the message $M$ sent by the encoder to the decoder. We
partition the set ${\cal I}^{c}$ into two sets, namely, the set ${\cal F}$
that identifies the \textquotedbl{}frozen\textquotedbl{} symbols $u_{{\cal F}}$
and the set ${\cal D}$ that identifies the \textquotedbl{}computable\textquotedbl{}
symbols $u_{{\cal D}}$. These sets are defined as
\begin{align}
\mathcal{F} & \triangleq\left\{ i\in[1:n]:Z(U_{i}|U^{i-1},X^{n},Y^{n})\geq1-2^{-n^{\beta}}\right\} \label{eq:setF}\\
\text{and}\ \mathcal{D} & \triangleq\left\{ i\in[1:n]:Z(U_{i}|U^{i-1})\leq2^{-n^{\beta}}\right\} ,\label{eq:setD}
\end{align}
where $\beta<\frac{1}{2}$ is a parameter of the underlying polar
coding scheme. Further, the source Bhattacharyya parameter $Z$ for
two random variables $A\in\{0,1,\dots,q-1\}$ and $B\in\mathcal{B}$
is defined as
\begin{equation}
Z(A|B)\triangleq\frac{1}{q-1}\sum_{\substack{a,a'\in{\cal A}:\\
a\neq a'
}
}\,\sum_{b\in\mathcal{B}}\sqrt{P_{A,B}(a,b)P_{A,B}(a',b)}.
\end{equation}
The Bhattacharyya parameters in \eqref{eq:setF} and \eqref{eq:setD}
are calculated based on the joint distribution $P(x^{n},y^{n},\hat{x}^{n},u^{n})$
given in (6). From \cite[Theorem 1]{HY13} and \cite[Theorem 4.3]{Sas12}, the
size $nR$ of the set ${\cal I}=\{1,2,\dots,n\}\setminus({\cal F}\cup{\cal D})$
is such that the rate $R$ is arbitrarily close to $R^{*}$ as $n$
grows large.

To determine the vector $u^{n}$, the following randomized successive
encoding rule is used for $i=1,2,\dots,n$:
\begin{equation}
u_{i}=\begin{cases}
\begin{array}{llc}
\hspace{-1ex}u_{i}\in{\cal U} & \text{with\,\ probability}\, P(u_{i}|u^{i-1},x^{n},y^{n})\ \text{if}\, i\in\mathcal{I},\\
\hspace{-1ex}u_{i}\in{\cal U} & \text{with\,\ probability}\, P(u_{i}|u^{i-1})\ \text{if}\, i\in\mathcal{D},
\end{array}\end{cases}\label{eq:encoding}
\end{equation}
where the probabilities in (10) are obtained from (6). The symbols
$u_{{\cal F}}$ are predetermined and are available at the decoder
prior to encoding. The vector $u_{\mathcal{I}}$ is sent to the decoder,
while the decoder obtains the vector $u_{{\cal D}}$ according to
a maximum likelihood rule as in \cite{MHSU14,ELKLJB14}:
\begin{equation}
\hat{u}_{i}=\begin{cases}
u_{i} & \text{for}\ i\in\mathcal{I},\\
f_{i}(\hat{u}^{i-1})\triangleq\arg\max_{u\in{\cal U}}P(u|\hat{u}^{i-1}) & \text{for}\ i\in\mathcal{D},\\
u_{i} & \text{for}\ i\in\mathcal{F}.
\end{cases}\label{eq:decoding}
\end{equation}
Finally, the codeword $\hat{x}^{n}$ is evaluated as $\hat{x}^{n}=G_{n}\hat{u}^{n}$.
\begin{rem}
Note that the decoding rule (11) does not require encoder and decoder
to share the set of Boolean functions needed by the scheme in \cite{HY13}
(see also \cite{MHSU14,ELKLJB14}), hence significantly simplifying
the implementation.
\end{rem}
\begin{rem}
If $\hat{X}^{n}$ is i.i.d. uniformly distributed in $\hat{{\cal X}}^{n}$
under (6), it follows from \cite{Ari10,HY13} that the set ${\cal D}$
has negligible size as $n$ grows large and hence the encoding and
decoding rules \eqref{eq:encoding} and \eqref{eq:decoding} can be
simplified by setting ${\cal D}=\emptyset$ as done in \cite{KU10}.
This condition applies, for instance, to the doubly symmetric binary
source studied in \cite{Yam83} (see Remark 2). 
Moreover, the encoding rule \eqref{eq:encoding} with ${\cal D}=\emptyset$
entails that the set of codewords $\hat{X}^{n}$ consists of the (approximately)
$q^{nR}$ sequences of a block coset code defined by the generator
matrix $G_{n}$ and by the frozen symbols $u_{{\cal F}}$. 
\end{rem}

\subsection{Optimality of polar codes}

In this section, we establish the optimality of polar codes for the
problem at hand. We start with the following proposition that entails
randomization over the frozen bits. The need for randomization is
removed in Proposition 2.%
\footnote{The notation $f(n)=O(g(n))$ means that there exist constants $n_{0}$
and $c$ such that for all integers $n>n_{0}$ the inequality $\left|f(n)\right|\leq c\left|g(n)\right|$
holds.%
}
\begin{prop}
Fix a triple $(R^{*}=I(XY;\hat{X}),D^{*}=\mathbb{E}(d(X,\hat{X})),\Delta^{*}=H(Y|\hat{X}))$
achieved by a joint distribution $P(x,y,\hat{x})$ in the rate-distortion-equivocation
region ${\cal R}^{*}$. For any $0<\beta'<\beta<\frac{1}{2}$, any
$\epsilon>0$, and for sufficiently large $n$, the sequence of rates
$R_{n}=\frac{1}{n}\left|\mathcal{I}\right|$, distortions $D_{n}=\frac{1}{n}\mathbb{E}(d^{n}(X^{n},\hat{X}^{n}))$,
and equivocations $\Delta_{n}=\frac{1}{n}H(Y^{n}|U_{{\cal I}})$ that
satisfy \begin{subequations}
\begin{align}
R_{n} & \leq R^{*}+\epsilon,\label{eq:prop1R}\\
D_{n} & \leq D^{*}+O\bigl(2^{-n^{\beta'}}\bigr),\label{eq:prop1D}\\
\Delta_{n} & \geq\Delta^{*}-O\bigl(2^{-n^{\beta'}}\bigr)\label{eq:prop1Delta}
\end{align}
\end{subequations} is achievable by the polar coding scheme (10)-(11), where the distortion $D_{n}$ and
the equivocation $\Delta_{n}$ are averaged over uniformly distributed frozen symbols
$u_{{\cal F}}$. 
\end{prop}
\begin{IEEEproof}
We first define the joint distribution induced by the encoding rule
\eqref{eq:encoding} under the assumption that the frozen symbols
are selected as i.i.d.~uniform variables with probability $\frac{1}{q}$
according to
\begin{multline}
P^{e}\left(x^{n},y^{n},u^{n},\hat{x}^{n}\right)=Q\left(x^{n},y^{n}\right)q^{-\left|\mathcal{F}\right|}\prod_{i\in\mathcal{D}}P\left(u_{i}|u^{i-1}\right)\\
\cdot\prod_{i\in\mathcal{I}}P\left(u_{i}|u^{i-1},x^{n},y^{n}\right)\cdot\boldsymbol{1}\{\hat{x}^{n}=u^{n}G_{n}\}.\label{eq:Pe}
\end{multline}
Note that in \eqref{eq:Pe} the codeword $\hat{X}^{n}$ is defined
based on the symbols $U^{n}$ selected by the encoder. We also introduce
the joint distribution that includes both \eqref{eq:encoding} and
the decoding rule in \eqref{eq:decoding} as
\begin{multline}
P^{d}\left(x^{n},y^{n},u^{n},\hat{u}^{n},\hat{x}^{n}\right)=Q\left(x^{n},y^{n}\right)q^{-\left|\mathcal{F}\right|}\prod_{i\in\mathcal{D}}P\left(u_{i}|u^{i-1}\right)\\
\cdot\prod_{i\in\mathcal{I}}P\left(u_{i}|u^{i-1},x^{n},y^{n}\right)\cdot\prod_{i\in\mathcal{F}\cup{\cal I}}\boldsymbol{1}\{\hat{u}_{i}=u_{i}\}\\
\cdot\prod_{i\in\mathcal{D}}\boldsymbol{1}\{\hat{u}_{i}=f_{i}(\hat{u}^{i-1})\}\cdot\boldsymbol{1}\{\hat{x}^{n}=\hat{u}^{n}G_{n}\}.\label{eq:Pd}
\end{multline}
The rate condition \eqref{eq:prop1R}
follows directly by extension of the arguments in \cite[Theorem 1]{HY13}
to alphabets of prime size and holds for any choice of the frozen
vectors. To prove \eqref{eq:prop1D} for the ensemble of codes inducing
the joint distribution by \eqref{eq:Pd}, we need to modify the arguments
in \cite{HY13} in order to account for possible decoding errors.
To this end, we define the probability of error as $P_{e}\triangleq\Pr_{P^{e}}[\hat{U}^{n}\neq U^{n}]$.\footnote{In the following, subscripts are used to identify the distribution with
  respect to which probabilities, expectations, and information measures are
  computed.}
Denoting
the decoding error event as $E\triangleq\{\hat{U}^{n}\neq U^{n}\}$, the distortion
$D_{n}(u_{{\cal F}})$ averaged over the frozen vectors
$u_{{\cal F}}$ satisfies
\begin{align}
\mathbb{E}_{P^{d}}[D_{n}(U_{{\cal F}})]= \hspace{1ex}& \frac{1}{n}\left((1-P_{e})\mathbb{E}_{P^{d}}[d(X^{n},\hat{X^{n}})\mid E^{c}]\right.\nonumber \\
 & + P_{e}\mathbb{E}_{P^{d}}[d(X^{n},\hat{X^{n}})|E]\Bigr)\nonumber \\
\leq \hspace{1ex} & \frac{1}{n}\left((1-P_{e})\mathbb{E}_{P^{d}}[d(X^{n},\hat{X^{n}})|E^{c}]\right.\nonumber \\
 & +P_{e}d_{\text{max}}\Bigr),\label{eq:EpD}
\end{align}
by the law of total probability and the boundedness of the distortion
metric. Moreover, we have
\begin{equation}
P^{d}\left(x^{n},y^{n},u^{n},\hat{u}^{n},\hat{x}^{n}|E^{c}\right)\!=\!\frac{P^{e}\left(x^{n},y^{n},u^{n},\hat{x}^{n}\right)\boldsymbol{1}\{u^{n}\!=\!\hat{u}^{n}\}}{1-P_{e}},
\end{equation}
and hence the first term in \eqref{eq:EpD} can be computed as
\begin{align}
\mathbb{E}_{P^{d}}\bigl[d(X^{n},\hat{X}^{n})|E^{c}\bigr]\notag\\
 & \hspace{-18ex}=\frac{1}{1-P_{e}}\cdot\sum_{x^{n},y^{n},u^{n},\hat{u}^{n},\hat{x}^{n}}P^{e}\left(x^{n},y^{n},u^{n},\hat{x}^{n}\right)\nonumber \\
 & \cdot\boldsymbol{1}\{u^{n}=\hat{u}^{n}\}d(x^{n},\hat{x}^{n})\nonumber \\
%
 & \hspace{-18ex}=\frac{1}{1-P_{e}}\mathbb{E}_{P^{e}}\bigl[d(X^{n},\hat{X}^{n})\bigr].\label{eq:EPDcond}
\end{align}
Furthermore, by \cite[Property 2]{SC13} we have the inequality
\begin{multline}
\frac{1}{n}\mathbb{E}_{P^{e}}\bigl[d(X^{n},\hat{X}^{n})\bigr]\leq\frac{1}{n}\mathbb{E}_{P}\bigl[d(X^{n},\hat{X}^{n})\bigr]+\frac{d_{\text{max}}}{n}\\
\cdot\bigl\Vert P_{X^{n},Y^{n},U^{n},\hat{X}^{n}}-P_{X^{n},Y^{n},U^{n},\hat{X}^{n}}^{e}\bigr\Vert,\label{eq:EPe}
\end{multline}
where $\bigl\Vert\cdot\bigr\Vert$ denotes the variational distance
of two distributions, and, by construction, $\frac{1}{n}\mathbb{E}_{P}\bigl[d(X^{n},\hat{X}^{n})\bigr]=D^{*}$
holds true. The variational distance in \eqref{eq:EPe} can be characterized
by following similar steps as in \cite{HY13,ELKLJB14} as (see the
Appendix for a sketch)
\begin{equation}
\bigl\Vert P_{X^{n},Y^{n},\hat{X}^{n},U^{n}}-P_{X^{n},Y^{n},\hat{X}^{n},U^{n}}^{e}\bigl\Vert=O(2^{-n^{\beta'}})\label{eq:vardist}
\end{equation}
for any $\beta'<\beta$. Finally, we obtain the following bound on the
probability of decoding error
\begin{equation}
P_{e}\overset{(a)}{\leq}\sum_{i\in{\cal D}}Z(U_{i}|U^{i-1})\overset{(b)}{\leq}\left|{\cal D}\right|2^{-n^{\beta}}\overset{(c)}{=}O(n2^{-n^{\beta}}),\label{eq:errorbound}
\end{equation}
where (a) follows from \cite[Proposition 2]{Ari09}, (b) is a consequence
of the definition of the set ${\cal D}$ in (8), and (c) follows
by noting that the cardinality of the set ${\cal D}$ is at most linear
in $n$. Using \eqref{eq:EpD}, along with \eqref{eq:EPDcond}-\eqref{eq:errorbound},
we have
\begin{align}
\mathbb{E}_{P^{d}}[D_{n}(U_{{\cal F}})] \leq &\ 
  \frac{1}{n}\bigl(\mathbb{E}_{P^{e}}[d(X^{n},\hat{X^{n}})]+P_{e}d_{\text{max}}\bigr),\nonumber \\
 \leq &\ \frac{1}{n}\bigl[nD^{*}+d_{\text{max}}(P_{e}\nonumber \\
 & \ +\parallel P_{X^{n},Y^{n},\hat{X}^{n},U^{n}}-P_{X^{n},Y^{n},\hat{X}^{n},U^{n}}^{e}\parallel)\bigr],\nonumber \\
 =& \ D^{*}+O(2^{-n^{\beta'}}),
\end{align}
which allows us to conclude that the distortion inequality \eqref{eq:prop1D}
is satisfied on average over the choice of the frozen vectors.

To prove \eqref{eq:prop1Delta}, we first observe that, by construction,
we have $\frac{1}{n}H_{P}(Y^{n}|\hat{X}^{n})=\Delta^{*}$. The achievable
average equivocation satisfies the equality
\begin{align}
\mathbb{E}_{P^{d}}[\Delta_{n}(U_{{\cal F}})] & =\frac{1}{n}H_{P^{d}}(Y^{n}|U_{{\cal I}},U_{{\cal F}})\nonumber \\
 & =\frac{1}{n}H_{P^{e}}(Y^{n}|U_{{\cal I}},U_{{\cal F}})=\mathbb{E}_{P^{e}}[\Delta_{n}(U_{{\cal F}})],\label{eq:EPd_Delta}
\end{align}
and further we have
\begin{align}
\mathbb{E}_{P^{e}}[\Delta_{n}(U_{{\cal F}})] & =\frac{1}{n}H_{P^{e}}(Y^{n}|U_{{\cal I}},U_{{\cal F}})\nonumber \\
 & \geq\frac{1}{n}H_{P^{e}}(Y^{n}|U^{n})=\frac{1}{n}H_{P^{e}}(Y^{n}|\hat{X}^{n}),\label{eq:EPe_Delta}
\end{align}
where the inequality in \eqref{eq:EPe_Delta} holds since  conditioning reduces
entropy, and the subsequent  equality holds due to  the one-to-one correspondence
between $\hat{X}^{n}$ and $U^{n}$ under $P^{e}$, respectively.


Using both the chain rule and the triangle inequality, we obtain
\begin{multline}
\bigl|H_{P}(Y^{n}|\hat{X}^{n})-H_{P^{e}}(Y^{n}|\hat{X}^{n})\bigr|\leq\\
\bigl|H_{P}(Y^{n},\hat{X}^{n})-H_{P^{e}}(Y^{n},\hat{X}^{n})\bigr|\\
+\bigl|H_{P}(\hat{X}^{n})-H_{P^{e}}(\hat{X}^{n})\bigr|.
\end{multline}
Now, by considering
\begin{multline}
\left\Vert P_{X}-Q_{X}\right\Vert =\sum_{x}\Bigl|\sum_{y}P(x,y)-Q(x,y)\Bigr|\\
\leq\sum_{x,y}\left|P(x,y)-Q(x,y)\right|=\left\Vert P_{X,Y}-Q_{X,Y}\right\Vert \label{eq:chainH}
\end{multline}
and by applying \cite[Lemma 2.7]{CK81} with \eqref{eq:vardist} and (25), we 
finally  obtain the bound
\[
\bigl|H_{P}(Y^{n}|\hat{X}^{n})-H_{P^{e}}(Y^{n}|\hat{X}^{n})\bigr|\leq O(n^{\beta'}2^{n^{\beta'}}).
\]
This shows that \eqref{eq:prop1Delta} is satisfied on average over
the choice of the frozen vectors.
\end{IEEEproof}
We now show that averaging over all frozen vectors is not required
to achieve the region $(R^{*},D^{*},\Delta^{*})$.
\begin{prop}
Any tuple $(R^{*},D^{*},\Delta^{*})$ in \eqref{eq:prop1R}, \eqref{eq:prop1D},
and \eqref{eq:prop1Delta} is achievable by time sharing between at
most two polar coding schemes defined by \eqref{eq:encoding} and
\eqref{eq:decoding} with different sequences of frozen symbols $u_{{\cal F}}$.
\end{prop}
\begin{IEEEproof}
We prove this statement by contradiction. To elaborate, if a sequence
of frozen vectors $u_{{\cal F}}$ exists such that for any fixed $\epsilon>0$
both conditions \eqref{eq:prop1D} and \eqref{eq:prop1Delta} are
satisfied, namely $D_{n}(u_{{\cal F}})\leq D^{*}+\epsilon$ and $\Delta_{n}(u_{{\cal F}})\geq\Delta^{*}-\epsilon$,
then the proof is complete. Now, we assume that none of the vectors
$u_{{\cal F}}$ satisfies both conditions. By the discussion above,
we can find a sufficiently large $n_{0}$ such that $\mathbb{E}_{P^{d}}[D_{n}(U_{{\cal F}})]\leq D^{*}+\epsilon$
and $\mathbb{E}_{P^{d}}[\Delta_{n}(U_{{\cal F}})]\geq\Delta^{*}-\epsilon$
for all $n\geq n_{0}$. Consider a coordinate system with origin at
$(D^{*}+\epsilon,\Delta^{*}-\epsilon)$ in the distortion-equivocation
plane (see Fig. 2). By assumption, for none of the vectors $u_{{\cal F}}$
the point $(D_{n}(u_{{\cal F}}),\Delta_{n}(u_{{\cal F}}))$ is in
the second (upper left) quadrant, while the average $(\mathbb{E}_{P^{d}}[D_{n}(U_{{\cal F}})],\mathbb{E}_{P^{d}}[\Delta_{n}(U_{{\cal F}})])$
lies in the second quadrant. Moreover, the average is in the convex
hull of the points $(D_{n}(u_{{\cal F}}),\Delta_{n}(u_{{\cal F}}))$,
which is a polytope. By simple geometric arguments, one of the edges
of this polytope must cross the second quadrant. Therefore, if the
vertices of this crossing edge are denoted as $(D_{n}(u_{{\cal F}1}),\Delta_{n}(u_{{\cal F}1}))$
and $(D_{n}(u_{{\cal F}2}),\Delta_{n}(u_{{\cal F}2}))$, then we can
find $0\leq\alpha\leq1$ such that $D^{\dagger}=\alpha D_{n}(u_{{\cal F}1})+(1-\alpha)D_{n}(u_{{\cal F}2})$,
and $\Delta^{\dagger}=\alpha\Delta_{n}(u_{{\cal F}1})+(1-\alpha)\Delta_{n}(u_{{\cal F}2})$,
and $(D^{\dagger},\Delta^{\dagger})$ lies in the second quadrant, hence
completing the proof.
\end{IEEEproof}
\begin{figure}[H]
\vspace{-1.3ex}
\centering\includegraphics[scale=0.43]{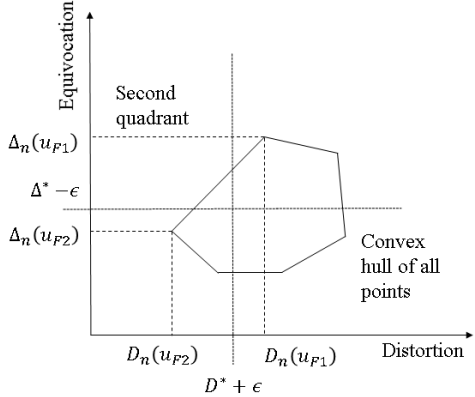} \vspace{-1ex}
 \protect\caption{Convex hull of points in the  equivocation-distortion plane.}

\vspace{-1ex}

\end{figure}

\begin{rem}
For the important case of the doubly symmetric source and Hamming
distortion, time sharing is not necessary. Hence, there exists a single
polar coding scheme defined by \eqref{eq:encoding} and \eqref{eq:decoding}
with a specific choice for the sequence of frozen bits $u_{{\cal F}}$
(and ${\cal D}=\emptyset$, see Remark 4) that achieves the desired
point $(R^{*},D^{*},\Delta^{*})$. This
can be seen from the fact that for each vector $u_{{\cal F}}$ we
have $\mathbb{E}_{P^{d}}[D_{n}(U_{{\cal F}})]=D_{n}(u_{{\cal F}})$ \cite{KU10}.
Now, since we know that $\mathbb{E}_{P^{d}}[\Delta_{n}(U_{{\cal F}})]\geq\Delta^{*}-\epsilon$,
there must be at least one frozen vector $u_{{\cal F}}$ such that
$\Delta_{n}(u_{{\cal F}})\geq\Delta^{*}-\epsilon$, which completes
the proof.
\end{rem}
\bibliographystyle{IEEEtran}
\bibliography{ITW}

\begin{thebibliography}{10}
\providecommand{\url}[1]{#1}
\csname url@samestyle\endcsname
\providecommand{\newblock}{\relax}
\providecommand{\bibinfo}[2]{#2}
\providecommand{\BIBentrySTDinterwordspacing}{\spaceskip=0pt\relax}
\providecommand{\BIBentryALTinterwordstretchfactor}{4}
\providecommand{\BIBentryALTinterwordspacing}{\spaceskip=\fontdimen2\font plus
\BIBentryALTinterwordstretchfactor\fontdimen3\font minus
  \fontdimen4\font\relax}
\providecommand{\BIBforeignlanguage}[2]{{%
\expandafter\ifx\csname l@#1\endcsname\relax
\typeout{** WARNING: IEEEtran.bst: No hyphenation pattern has been}%
\typeout{** loaded for the language `#1'. Using the pattern for}%
\typeout{** the default language instead.}%
\else
\language=\csname l@#1\endcsname
\fi
#2}}
\providecommand{\BIBdecl}{\relax}
\BIBdecl

\bibitem{STRVD13}
L.~Sankar, W.~Trappe, K.~Ramchandran, H.~V. Poor, and M.~Debbah, ``The role of
  signal processing in meeting privacy challenges: {A}n overview,'' \emph{IEEE
  Signal Processing Magazine}, vol.~30, pp. 95--1096, Sep. 2013.

\bibitem{Yam83}
H.~Yamamoto, ``A source coding problem for sources with additional outputs to
  keep secret from the receiver or wiretapper,'' \emph{IEEE
  Trans.~Inf.~Theory}, vol. IT-29, no.~6, Nov. 1983.

\bibitem{AS00}
R.~Agrawal and R.~Srikant, ``Privacy-preserving data mining,'' \emph{ACM Sigmod
  Record}, vol.~29, no.~2, pp. 439--450, Jun. 2000.

\bibitem{DSNS08}
C.~Dwork, F.~McSherry, K.~Nissim, and A.~Smith, ``Calibrating noise to
  sensitivity in private data analysis,'' in \emph{Lecture Notes in Computer
  Science}.\hskip 1em plus 0.5em minus 0.4em\relax Springer, 2006, vol. 3876,
  pp. 265--284.

\bibitem{SRP13}
L.~Sankar, S.~R. Rajagopalan, and H.~V. Poor, ``Utility-privacy tradeoffs in
  databases: An information-theoretic approach,'' \emph{IEEE
  Trans.~Inf.~Forensics and Security}, vol.~8, no.~6, pp. 838--852, Jun. 2014.

\bibitem{MSFM14}
A.~Makhdoumi, S.~Salamatian, N.~Fawaz, and M.~M{\'e}dard, ``From the
  information bottleneck to the privacy funnel,'' in \emph{Proc.~IEEE
  Inf.~Theory Workshop}, Hobart, Australia, Nov. 2014, pp. 502--506.

\bibitem{KOS14}
K.~Kittichokechai, T.~J. Oechtering, and M.~Skoglund, ``Lossy source coding
  with reconstruction privacy,'' in \emph{Proc.~IEEE Int.~Symposium on
  Inform.~Theory}, Honululu, HI, Jul. 2014, pp. 386--390.

\bibitem{Ari09}
E.~Arikan, ``Channel polarization: {A} method for constructing
  capacity-achieving codes for symmetric binary-input memoryless channels,''
  \emph{IEEE Trans.~Inf.~Theory}, vol.~55, no.~7, pp. 3051--3073, Jul. 2009.

\bibitem{HY13}
J.~Honda and H.~Yamamoto, ``Polar coding without alphabet extension for
  asymmetric models,'' \emph{IEEE Trans.~Inf.~Theory}, vol.~59, no.~12, pp.
  7829--7838, Dec. 2013.

\bibitem{MHSU14}
M.~Mondelli, S.~H. Hassani, I.~Sason, and R.~Urbanke, ``Achieving {Marton's}
  region for broadcast channels using polar codes,'' [Online]. Available at
  http://arxiv.org/abs/1401.6060, Jan. 2014.

\bibitem{Sas12}
E.~Sasoglu, ``Polarization and polar codes,'' in \emph{Foundations and Trends
  in Communications and Information Theory}.\hskip 1em plus 0.5em minus
  0.4em\relax NOW Publishers, 2012, vol.~8, no.~4, pp. 259--381.

\bibitem{KU10}
S.~B. Korada and R.~Urbanke, ``Polar codes are optimal for lossy source
  coding,'' \emph{IEEE Trans.~Inf.~Theory}, vol.~56, no.~4, pp. 1751--1768,
  Apr. 2010.

\bibitem{ELKLJB14}
E.~E. Gad, Y.~Li, J.~Kliewer, M.~Langberg, A.~Jiang, and J.~Bruck, ``Asymmetric
  error correction and flash-memory rewriting using polar codes,'' [Online].
  Available at http://arxiv.org/abs/1410.3542, Oct. 2014.

\bibitem{Ari10}
E.~Arikan, ``Source polarization,'' in \emph{Proc.~IEEE Int.~Symposium on
  Inform.~Theory}, Austin, TX, Jun. 2010, pp. 899--903.

\bibitem{SC13}
C.~Schieler and P.~Cuff, ``Rate-distortion theory for secrecy systems,''
  [Online]. Available at http://arxiv.org/abs/1305.3905, May 2013.

\bibitem{CK81}
I.~Csiszar and J.~K{\"o}rner, \emph{Information Theory: Coding Theorems for
  Discrete Memoryless Systems}.\hskip 1em plus 0.5em minus 0.4em\relax New
  York: Academic Press, 1981.

\bibitem{GAG13}
N.~Goela, E.~Abbe, and M.~Gastpar, ``Polar codes for broadcast channels,''
  [Online]. Available at http://arxiv.org/abs/1301.3905, Jan. 2013.

\end{thebibliography}

\vspace{1ex}
\appendix
\vspace{1ex}
Proof sketch of \eqref{eq:vardist}: {\small{}
\begin{align*}
\left\Vert P_{X^{n},Y^{n},U^{n},\hat{X}^{n}}-P_{X^{n},Y^{n},U^{n},\hat{X}^{n}}^{e}\right\Vert \\
 & \hspace{-33ex}=\sum_{u^{n},x^{n},y^{n}}\left|P(u^{n},x^{n},y^{n})-P^{e}(u^{n},x^{n},y^{n})\right|\\
 & \hspace{-33ex}\overset{(a)}{=}\sum_{u^{n},x^{n},y^{n}}\biggl|\sum_{i=1}^{n}[P(u_{i}|u^{i-1},x^{n},y^{n})-P^{e}(u_{i}|u^{i-1},x^{n},y^{n})]\cdot\\
 & \hspace{-30ex}Q(x^{n},y^{n})\prod_{j=1}^{i-1}P(u_{j}|u^{j-1},x^{n},y^{n})\prod_{j'=i+1}^{n}P^{e}(u_{j'}|u^{j'-1},x^{n},y^{n})\biggr|\\
 & \hspace{-33ex}\overset{(b)}{\leq}\sum_{i\in{\cal F}}\sum_{u^{i},x^{n},y^{n}}\left|P(u_{i}|u^{i-1},x^{n},y^{n})-P^{e}(u_{i}|u^{i-1},x^{n},y^{n})\right|\cdot\\
 & \hspace{-30ex}Q(x^{n},y^{n})\prod_{j=1}^{i-1}P(u_{j}|u^{j-1},x^{n},y^{n})\\
 & \hspace{-33ex}=\sum_{i\in{\cal F}}\sum_{u^{i-1},x^{n},y^{n}}\hspace{-2ex}P(u^{i-1},x^{n},y^{n})\left\Vert P_{U_{i}|u^{i-1},x^{n},y^{n}}\!-\! P_{U_{i}|u^{i-1},x^{n},y^{n}}^{e}\right\Vert \\
 & \hspace{-33ex}\overset{(c)}{\leq}\sum_{i\in{\cal F}}\sum_{u^{i-1},x^{n},y^{n}}P(u^{i-1},x^{n},y^{n})\cdot\\
 & \hspace{-30ex}\sqrt{(2\ln2)D(P_{U_{i}|u^{i-1},x^{n},y^{n}}\parallel P_{U_{i}|u^{i-1},x^{n},y^{n}}^{e})}\\
 & \hspace{-33ex}=\sum_{i\in{\cal F}}\sqrt{(2\ln2)D(P_{U_{i}}\parallel P_{U_{i}}^{e}|U^{i-1},X^{n},Y^{n})}\\
 & \hspace{-33ex}\overset{(d)}{=}\sum_{i\in{\cal F}}\sqrt{(2\ln2)(1-H_{P}(U_{i}|U^{i-1},X^{n},Y^{n}))}\\
 & \hspace{-33ex}\overset{(e)}{\leq}\sum_{i\in{\cal F}}\sqrt{(2\ln2)(1-(Z(U_{i}|U^{i-1},X^{n},Y^{n}))^{2})}\\
 & \hspace{-33ex}\overset{(f)}{\leq}n\sqrt{(4\ln2)2^{-n^{\beta}}}=O(2^{-n^{\beta'}})
\end{align*}
}Here, the equalities and inequalities follow from (a) a telescopic
expansion, (b) the fact that the distributions $P$ and $P^{e}$ are the same for $i\notin{\cal F}$,
(c) Pinsker's inequality 
where
$D(\cdot\Vert\cdot)$ is the relative entropy, (d) \cite[Lemma 10]{GAG13},
(e) \cite[Proposition 4.8]{Sas12}, (f) \eqref{eq:setF}.
\end{document}